\documentclass[
12pt,
3p,
review,
]{elsarticle}
\usepackage{hyperref}
\usepackage{graphicx,graphics}
\usepackage{placeins}
\usepackage{amsmath}
\usepackage{amssymb}
\usepackage{mathtools}
\usepackage{array}
\usepackage{hyperref}
\usepackage{color}
\usepackage{soul}
\usepackage{dcolumn}
\usepackage{bm}
\usepackage{todonotes}
\usepackage[utf8]{inputenc}
\usepackage[english]{babel}
\usepackage{libertine}
\usepackage{pgfplots}
\usepgfplotslibrary{polar}
\pgfplotsset{compat=1.9}
\usetikzlibrary{calc}

\usepackage[title]{appendix}

\usepackage{multirow}

\usepackage{layouts}

\usepackage{caption}
\usepackage{subcaption}

\renewcommand{\l}{\mathopen{}\mathclose\bgroup\left}
\renewcommand{\r}{\aftergroup\egroup\right}

\let\originaleps=\epsilon
\let\epsilon=\varepsilon
\let\varepsilon=\originaleps
\usepackage{stmaryrd}

\mathchardef\hy="2D

\definecolor{mydark_blue}{RGB}{0, 0, 139}
\definecolor{myblue}{RGB}{0, 0, 255}
\definecolor{mycyan}{RGB}{0, 255, 255}  
\definecolor{mygreen}{RGB}{0, 255, 0}
\definecolor{myyellow}{RGB}{255, 255, 0}
\definecolor{myred}{RGB}{255, 0, 0}
\definecolor{mydark_red}{RGB}{139, 0, 0}
\definecolor{myblack}{RGB}{0, 0, 0}

\definecolor{BRY_1}{RGB}{  0,  0,255}
\definecolor{BRY_2}{RGB}{127,  0,127}
\definecolor{BRY_3}{RGB}{255,  0,  0}
\definecolor{BRY_4}{RGB}{255,127,  0}
\definecolor{BRY_5}{RGB}{255,255, 85}

\journal{XXXXXXX}

\begin{document}

\begin{frontmatter}
\title{The inadequacy of the geometric features of MA islands in relating bainite microstructures to composition and processing conditions}

\author[mymainaddress]{Vinod Kumar\fnref{fn1}}
\author[mymainaddress]{Sharukh Hussain\fnref{fn1}}
\author[mysecondaryaddress]{Priyanka S}
\author[mymainaddress]{P G Kubendran Amos\corref{mycorrespondingauthor}}
\ead{prince@nitt.edu}

\cortext[mycorrespondingauthor]{P G Kubendran Amos}

\fntext[fn1]{The authors contributed equally.}

\address[mymainaddress]{Theoretical Metallurgy Group,
Department of Metallurgical and Materials Engineering,\\
National Institute of Technology Tiruchirappalli, \\
Tamil Nadu, India}

\address[mysecondaryaddress]{School of Electronics Engineering,
VIT-AP University,\\
Amaravati, Andra Pradesh, India
}

\begin{abstract} 

Achieving desired properties in bainite steels with MA islands demands understanding the affect of processing conditions and composition on their size and morphology. 
Generally, this understanding is gained by studying the change in the size and morphology of the MA islands with composition and the processing conditions. 
In the present work, around 8500 MA islands dispersed across of approximately 1500 bainite microstructures are investigated to comprehend the influence of composition and heat treatment cycle on the geometric features. 
The geometric features considered in this study include polygon area metric, compactness and aspect ratio. 
A thorough statistical analysis of these features across bainite steels of different compositions and processing conditions unravel that, though there are minor changes, no characteristic variation is introduced in the size and morphology of the MA islands. 
In other words, the distribution of the various forms of MA islands are almost identical in bainite steels of different chemistry and heat treatment, thereby indicating the inadequacy of geometric features in explicating the affect of processing conditions on microstructure.

\end{abstract}

\begin{keyword}
Bainite microstructure, Martensite-Austenite Islands, Isothermal Transformation, Microstructure-Processing-Parameter Relation, Microstructure Analysis, Multiphase System
\end{keyword}

\end{frontmatter}


\section{Introduction}

Austenitic, ferritic, martensitic and bainitic steels have been around for a few decades now~\cite{krauss2015steels}.
Besides being used in wide-range of applications, these steels are progressively analysed to understand and enhance their typifying behaviour. 
In amongst these four steels, a comprehensive understanding of one is yet to be gained. This evasive ferrous alloy is the bainitic steel~\cite{fielding2013bainite}. 
One of the primary reasons for the lack of substantial understanding of bainite steel is its microstructure, which is significantly complex when compared to its counterparts~\cite{bhadeshia2001bainite,huang2014secondary}.
The characteristic appearance of bainite, on the microscopic scale, is intricate to the extent that it took years, and some debates, to gain a consensus on its definition~\cite{aaronson1990bainite}. 
Accordingly, the competitive growth of eutectoid phases, as opposed to cooperative growth, is deemed to yield bainite~\cite{lee1988mechanisms}. 
Stated otherwise, bainite microstructure reflects a characteristic distribution of ferrite and cementite resulting from non-cooperative growth.
Despite this general definition, there exists bainite microstructures with marginal or no carbides~\cite{tian2019transformation}.
Moreover, irrespective of carbide, microstructures of bainite steels are invariably associated with multiple other phases~\cite{bramfitt1990perspective}. 
Given the non-cooperative formation, these phases exhibit diverse morphologies. 
The combination of several phases and their manifold morphologies contribute to the complexity of bainite microstructures.

In spite of the possible presence of several phases, based on the distribution of ferrite and cementite, particularly carbide, two types of bainite are realised. 
These are referred to as upper and lower bainite~\cite{yin2017morphology}. 
When the finish cooling temperature (or holding temperature) is relatively high, carbides form between the individual ferrite structures, yielding upper bainite. 
Increased undercooling (or low holding temperature) results in lower bainite characterised by the presence of carbides within ferrite. 
The transformation temperature, by dictating the migration of carbon, is primarily responsible for these two types of bainite.
In other words, while enhanced carbon migration from ferrite forms upper bainite, restricted diffusion leads to lower bainite with intra-lath carbides. 
The classification of bainite into upper and lower, though elegant, is rather limited. 
This limitation stems from the almost exclusive consideration of isothermal transformation of bainite. 
For instance, carbides are observed both within and between the ferrite structures in bainite emerging from continuous cooling~\cite{krauss1995ferritic}. 
More importantly, the generally prevalent \textit{incomplete transformation}, under the additional influence of the alloying elements introduces secondary phases, besides ferrite and cementite~\cite{wu2017incomplete}.

During bainite transformation, the volume fraction of the phases rarely adheres to the lever rule. 
Generally, the phase fraction is noticeably less than the predicted proportion. 
This is referred to as the incomplete transformation~\cite{bhadeshia1982bainite,schoof2020influence}. 
The reason for the incomplete transformation varies with the competing schools of thought on bainite transformation.
Displacive theory, characterised by the diffusionless growth of bainite sheaves followed by the expulsion of carbon, attributes the incomplete transformation to decrease in the driving force caused by the accumulation of carbon~\cite{caballero2013new}. 
On the other hand, the solute drag introduced during the reconstructive formation of bainite is viewed as the reason for the deviation from the expected phase-fraction in the opposite school of thought~\cite{hillert1994diffusion}. 
Irrespective of the underlying reason, the incomplete transformation, by preserving the parent austenite, facilitates the introduction of varied phases including pearlite~\cite{caballero2009new}, martensite~\cite{lu2021effect} and retained austenite~\cite{wang2012new}. 
The phases which are introduced, besides ferrite and cementite, by the incomplete transformation, are referred to as the \textit{secondary phases}. 
The secondary phases can form alongside upper or lower bainite, resulting in degenerate bainite. Correspondingly, in addition to conventional upper and lower bainite, the incomplete transformation renders degenerate upper and lower bainite with secondary phases~\cite{zhang2016high}.

Phases introduced subsequently after the incomplete transformation are determined primarily by the cooling regime, including final cooling temperature, and alloying elements. 
Depending on these factors, the untransformed parent austenite evolves into one or more combination of secondary phases~\cite{zajac2005characterisation}. 
Slower cooling rate, or higher holding temperature, facilitates long-range migration of carbon, minimising any spatial gradient in concentration, particularly along the interface separating bainite-ferrite and austenite. 
The dissipation of carbon, along with suitable alloying elements, hinders the formation of carbides. Furthermore, increase in carbon content stabilises austenite by lowering the martensite-start temperature. This carbon-induced stabilisation results in the appearance of retained austenite. 
The long range diffusion at high temperature or low cooling rate cannot ensure a homogenised distribution of carbon in parent austenite.
Correspondingly, as the evolution continues, pockets of low carbon regions in austenite transform to martensite. 
These localised martensite phases surrounded, and intertwined, by relatively carbon-rich austenite form a secondary constituent called martensite-austenite (MA) island~\cite{zhang2002accurate}.

Martensite-austenite islands are not uncommon secondary constituents of bainite steel. 
And more importantly, given that the formation of MA islands depends largely on the distribution of carbon, their morphologies are generally complex~\cite{park2001interpretation}. 
During bainite transformation, carbon distribution is governed by cooling regime along and alloying elements, particularly those dictating the carbide formation~\cite{timokhina2001microstructure}. 
Correspondingly, the size, distribution and morphology of MA islands are effected by these parameters.
Stated otherwise, \textit{a bainite steel of definite composition, exposed to a specific cooling regime, encompasses varied forms of MA islands.}
A thoughtful consideration of the varied and intricate morphology of MA islands is necessary owing to their noticeable influence on mechanical behaviour of bainite steels.
From yield strength~\cite{isasti2014microstructural} to bendability~\cite{kaijalainen2016influence}, MA islands have been realised to affect a wide range of mechanical properties. 
Fracture toughness, in particular, is significantly altered by the features of MA islands beyond its volume-fraction~\cite{terasaki2015effect}. 
The interface separating MA island and the surrounding matrix, owing to its characteristic decohesion, is viewed as the potential site for crack nucleation~\cite{avramovic2009effect}.
Correspondingly, the shape of the MA island, by altering the crack-initiation stress, governs void formation~\cite{taboada2019substructure}. 
Besides formation, the propagation of a crack is dictated by its interaction with MA islands~\cite{mao2018relationship}. 
This interaction, which is mainly driven by the morphology and aspect ratio, regulates the resistance offered by bainite steels to the growth of crack, and determines its propagating path. 
Given the significance of MA islands, desired mechanical properties in bainite steels can only be established by suitably varying the associated morphology and aspect ratio.
Achieving such control over MA islands begins with relating the corresponding features with composition and processing parameters. 

In multiphase systems, phase-fractions are generally related to composition and processing parameters, owing to their principal focus on properties.
However, in bainite steels with MA islands, an exclusive consideration of volume fraction is hugely restrictive. 
Moreover, the intricate morphology of MA island limits the significance of conventional geometric factors, including perimeter, aspect ratio, area and compaction, in understanding the impact of governing parameters like composition and processing temperatures. 
Consequently, a convincing relation between the governing factors and MA islands can be only achieved when the intricacy of the varied shapes is given full consideration. 
Conventional approaches are rarely devised to capture such complex structures. 
Therefore, deep learning techniques are employed to understand and relate bainite steels with MA islands to the governing parameters and properties. 
Recently, by labelling different bainite steels based on their Charpy Impact Value (CIV), features of MA islands along with other components of microstructures have \textit{collectively} been related to these alloys through deep-learning based classification technique~\cite{ackermann2023explainable}. 

\section{Principal dataset}

In the present work, the dataset reported by Iren et al.~\cite{iren2021aachen} serves as the principal source of information on MA islands. 
For generating this dataset, steels of eight different chemical compositions have been considered. 
Bainite transformation is induced in these different steels at 400 and 500$^o$C. 
Morphology and distribution of the resulting phases are characterised by Scanning Electron Microscope in 4000x magnification. 
Human experts have been involved to identify and locate MA islands in bainite microstructures of different steels.
Besides the location of  MA islands indicated as points-of-interest, the morphology assumed of the islands is also realised by superimposing a polygon onto the structures. 
Both the position of MA islands and coordinates of the polygons, capturing its morphology, are recorded through \textit{Shapely}.

The dataset, in its entirety, includes 1705 high-quality SEM images. 
In these micrographs, 8909 points-of-interests representing MA islands have been realised by experts. 
Metadata is included in the work of Iren et al.~\cite{iren2021aachen}, in addition to the images, that encompasses coordinates describing the spatial positions of the MA island, and polygons reflecting its morphology. 
Furthermore, each micrograph is related to its corresponding steel with characteristic composition and processing temperature. 
The link between a composition,  processing temperature and micrograph is established through a dedicated URL in the metadata.
Parameters involved in capturing the microstructure, such as distance to the edge of the sample, tilt angle and photograph direction, are also included. 
Though images of the different steels are sufficiently distinguished through the categorical variables, specific composition is not disclosed. 
Therefore, in the present work, the designations assigned to the different steels in the metadata will be employed, and the configuration of MA islands will be discussed accordingly.

Working on a relatively limited set of data, existing works claim definitive effects of processing parameters and composition on geometric features of MA islands~\cite{caballero2012influence}. 
For instance, noticeable disparity is reported in the range (maximum and minimum) of perimeter and area of MA islands across two steels treated under different cooling regimes~\cite{ackermann2020effect}. 
This observation seemingly affirms the means of relating MA island to processing temperature and composition through similar geometric factors alone, without the need for any sophisticated deep-learning techniques. 
Considering that the metadata includes information on the geometric features of every MA islands, the preliminary investigation seeks to unravel any characteristic disparity in these features introduced by the processing temperature and composition. 

\section{Geometric features}

\subsection{Aspect ratio}

\begin{figure}
    \centering
    
    \begin{subfigure}[b]{1.0\textwidth}
        \includegraphics[width=\textwidth]{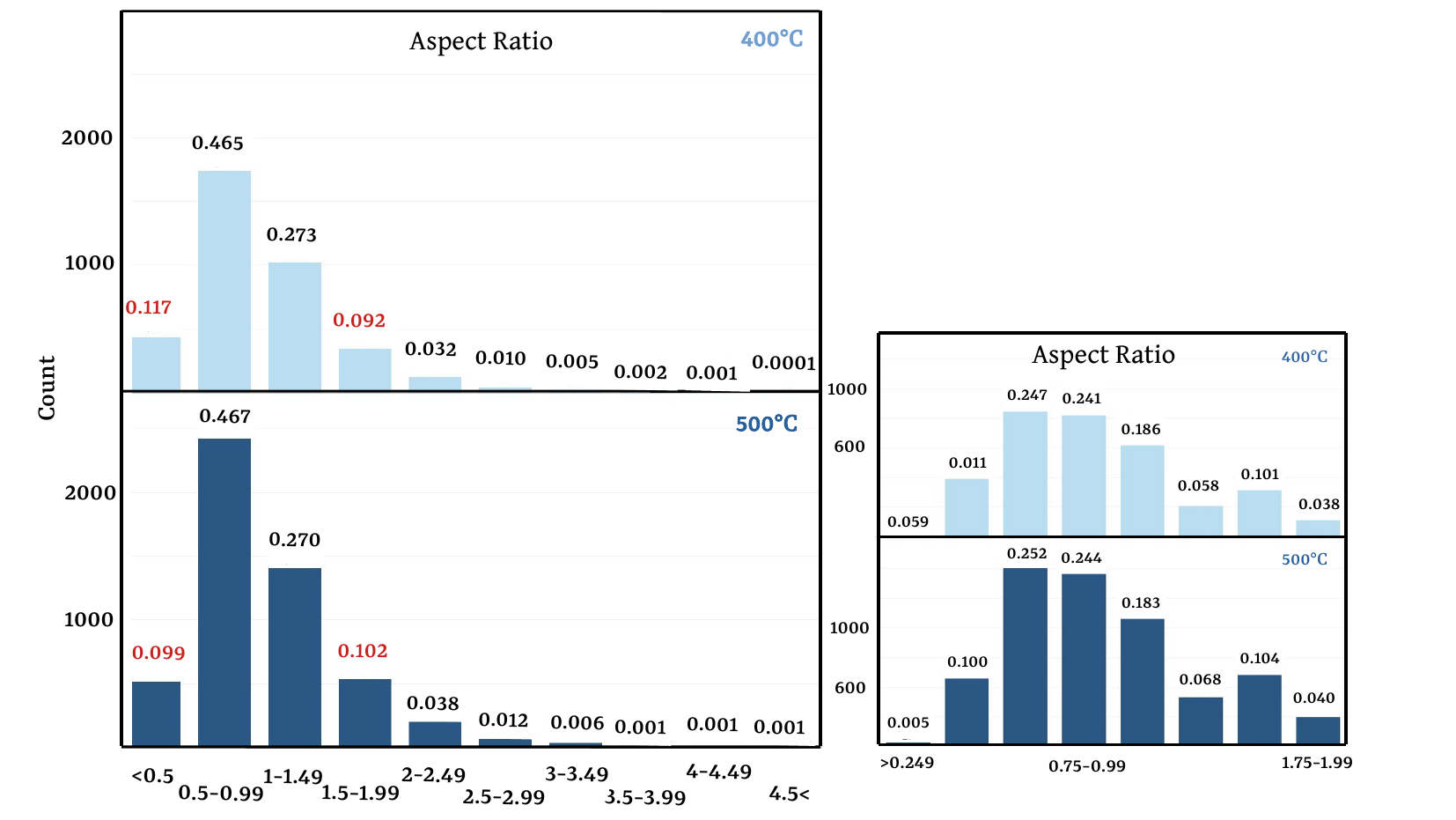}
        \caption{}
        \label{fig:AspectRatio}
    \end{subfigure}

    \begin{subfigure}[t]{1.0\textwidth}
    \centering
        \includegraphics[width=\textwidth]{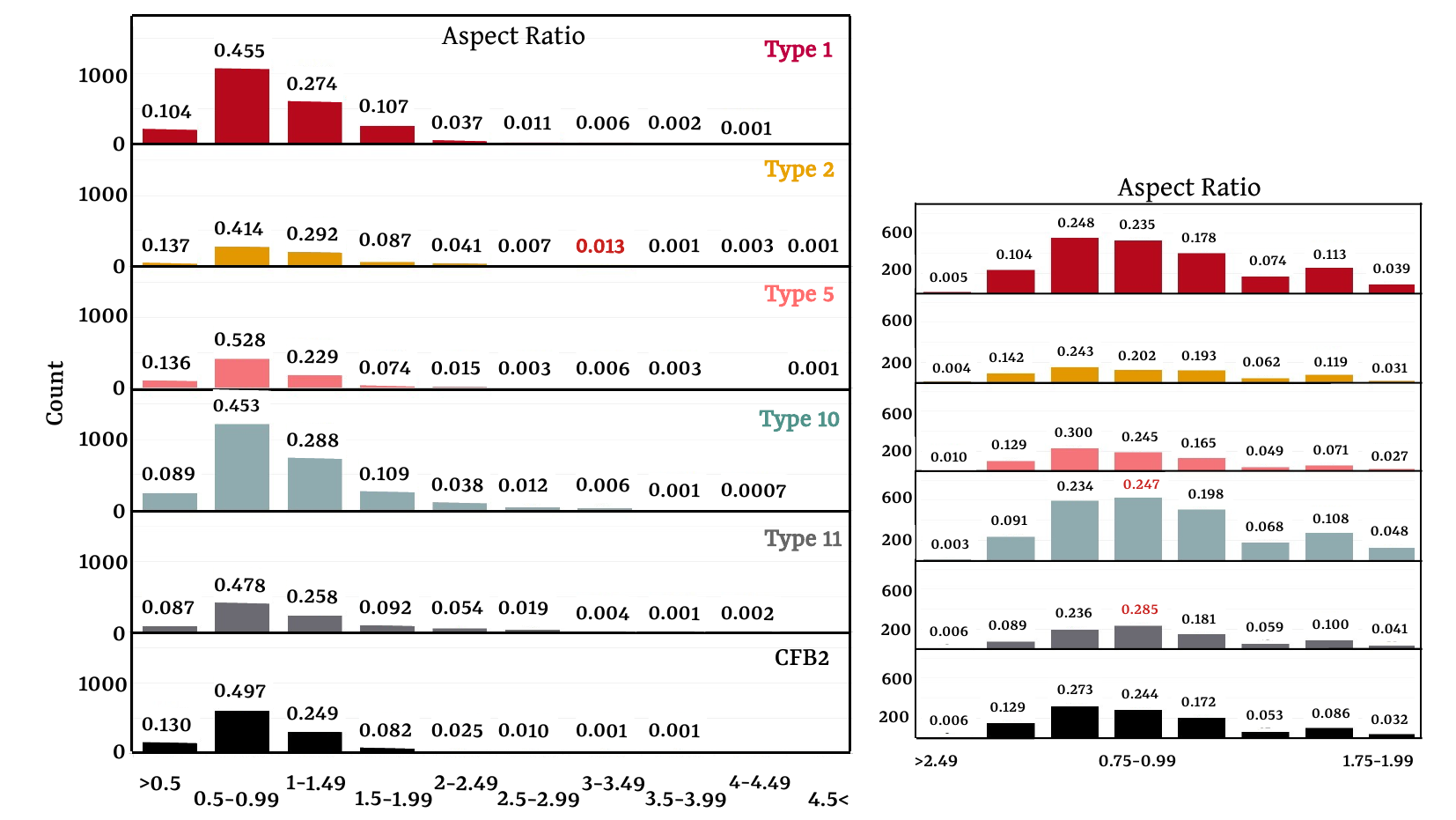}
        \caption{}
        \label{fig:Aspect_comp}
    \end{subfigure}
    
    \caption{Aspect-ratio based population distribution of MA islands in bainite microstructures (a) transformed at different temperatures and (b) pertaining to different steels. While the counts are represented along the axis, the corresponding number fraction is included within the histogram. The number fractions that are rather different across the temperature or composition are highlighted in red. }  
    \label{fig:}
\end{figure}

A geometric factor which gives a qualitative understanding of the span of a complex shape is aspect ratio. 
By encapsulating MA island in a bounding box, the corresponding aspect ratio is estimated by dividing the height with the width. 
This dimensionless parameter is ascertained for all MA island in different bainite microstructures and included in the metadata. 
Distribution of MA islands across the observed range of aspect ratios is illustrated in Fig.~\ref{fig:AspectRatio}. 
Distinction is made between MA islands formed at 400 and 500$^o$C through appropriate scheme. 
Both the count and number fraction of islands within a specific aspect ratio is included in the illustration. 

Fig.~\ref{fig:AspectRatio} unravels almost identical distribution in aspect ratios of MA islands across bainite microstructures formed at 400 and 500$^o$C. 
Stated otherwise, a disparity in the range or population density of aspect ratio is not introduced by the processing temperatures in the present corpus of data.
Irrespective of the transformation temperature, MA islands with aspect ratio ranging from 0.5 to 1.49 [0.5,1.5) is observed in all steels considered in this study. 
In spite of a analogous pattern, a marginal disparity in aspect ratio [1.5, 2.0) is visible in Fig.~\ref{fig:AspectRatio}.  
While number of MA islands with aspect ratio [1.5, 2.0) follow the predominant range [0.5, 1.5) in microstructures formed at 500$^o$C, it is interrupted by smaller island with less than 0.5 aspect ratio in steel processed at 400$^o$C. 
Apart from this minor difference, the entire trend in the distribution in aspect ratio of MA islands remains identical across the processing temperatures. 
Distribution in the densely populated range of aspect ratio is examined closely by reducing the original bin size. 
The enhanced view of this distribution, included in Fig.~\ref{fig:AspectRatio}, discloses no additional disparity except offering a refined view of the previously observed disparity. 

Similar to processing temperatures, aspect-ratio based population distribution of MA islands across different steels with characteristic composition is shown in Fig.~\ref{fig:Aspect_comp}. 
This illustration indicates identical trend in the distribution of MA islands over the varied ranges of aspect ratio independent of the composition. 
In all the bainite steels, MA islands are largely confined to the aspect ratio of 3.0, with the range upto 1.99 densely populated. 
Distinctly in type 2 steel, a fraction of MA islands, greater than the outliers (less than 1$\%$), occupy aspect ratio between [3.0, 3.5). 
A refined view of the densely populated section of the distribution in Fig.~\ref{fig:Aspect_comp} unravels further disparity. 

Fig.~\ref{fig:Aspect_comp} illustrates the distribution of MA islands across a refined ranges of aspect ratio. 
While the ranges between 0.5 and 1.25 are most densely populated, the maximum count is not restricted to one bin. 
Highest fractions of MA islands, in type 10 and 11, exhibit aspect ratio [0.75, 1.0), as opposed to other steels.  Put simply, in most steels, the aspect ratio [0.5, 0.75) exhibits maximum count except for type 10 and 11. However, the difference in counts across the two densely populated bins of type 10 and 11 steels is rather not significant. 
Apart from this marginal disparity, the overall pattern even in the refined illustration remains same irrespective of the composition. 

\subsection{Polygon Area Metric (PAM)}

\begin{figure}
    \centering
    
    \begin{subfigure}[b]{1.0\textwidth}
        \includegraphics[width=\textwidth]{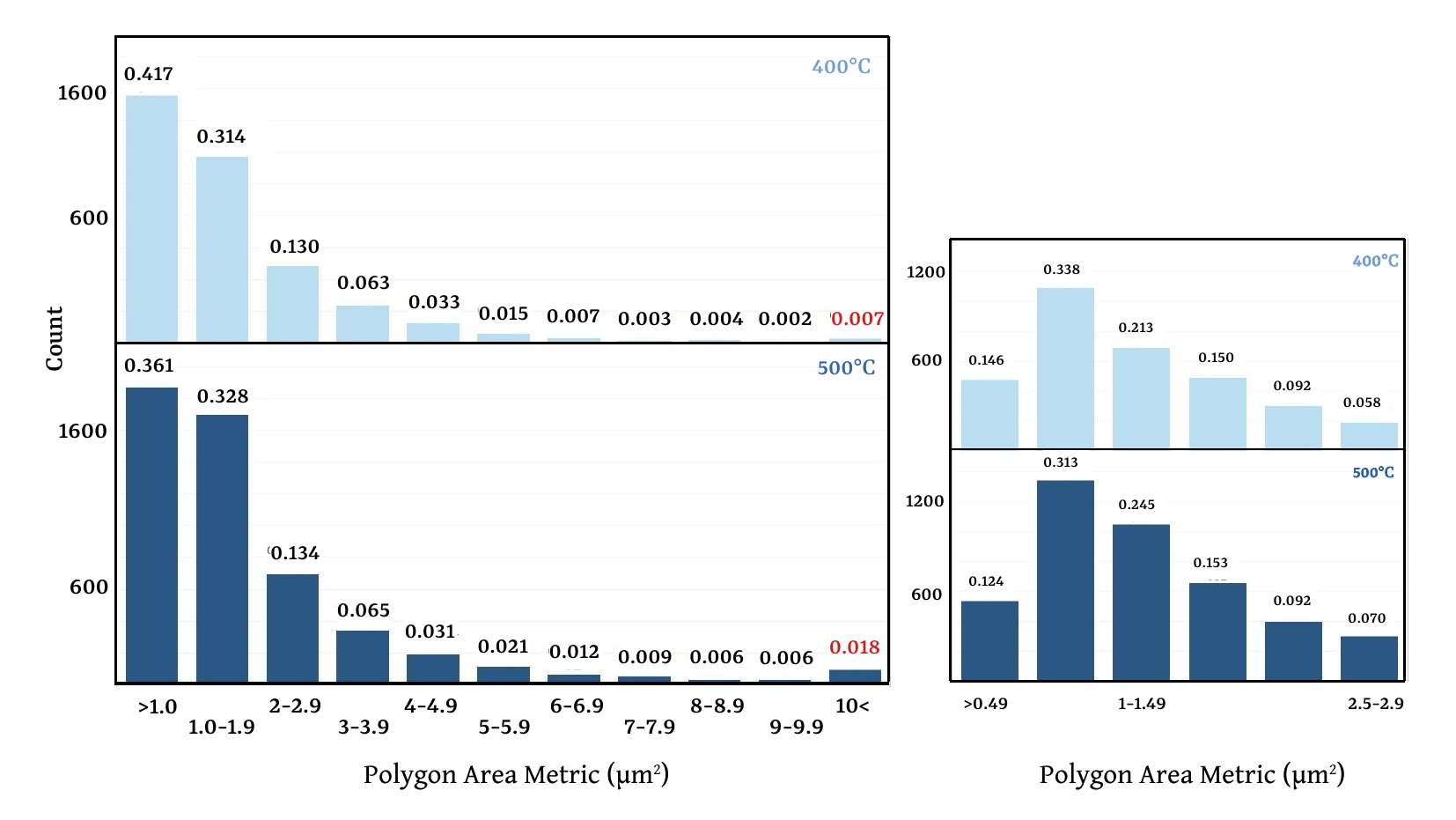}
        \caption{}
        \label{fig:PAM_temp}
    \end{subfigure}

    \begin{subfigure}[t]{1.0\textwidth}
    \centering
        \includegraphics[width=\textwidth]{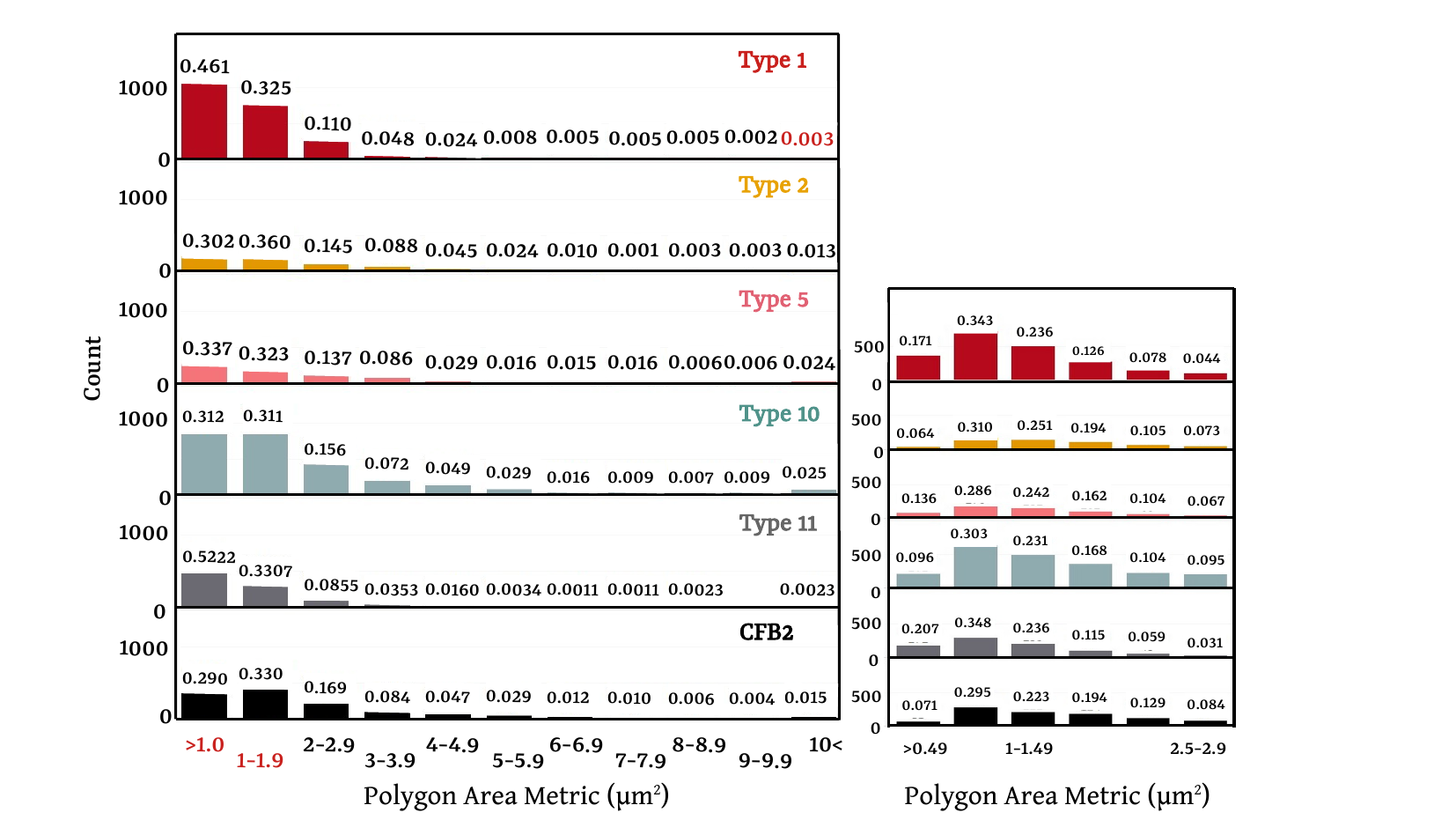}
        \caption{}
        \label{fig:PAM_comp}
    \end{subfigure}
    
    \caption{Polygon Area Metric (PAM) based population distribution of MA islands in bainite microstructures (a) transformed at different temperatures and (b) pertaining to different steels. While the counts are represented along the axis, the corresponding number fraction is included within the histogram. The number fractions (or PAM range) that are rather different across the temperature or composition are highlighted in red. }  
    \label{fig:}
\end{figure}

In addition to identifying and locating MA islands in bainite microstructures, their shapes are realised through appropriate coordinates.
These coordinates are subsequently used to build and calculate the area of the polygon that encapsulates MA island. 
Area of MA island expressed in micrometers is included in the metadata as polygon area metric (PAM). 
This area of MA islands, formed at different processing temperatures, are separately illustrated in Fig.~\ref{fig:PAM_temp}. 
More so than aspect ratio, the distribution of PAM is similar across the different transformation temperatures. 
Number of MA islands with increased area, in both temperatures,  gradually decreases with a minor raise in the count beyond 10$\mu$m$^2$. 
The degree of increase in the population of MA islands with area greater than 10$\mu$m$^2$ is slightly larger in bainite transformed at 500$^o$C when compared to 400$^o$C. 
This increased number density of MA islands is indicative of the enhanced carbon diffusion at higher transformation temperature~\cite{zajac2005characterisation}. 

One other marginal difference introduced by the transformation temperature is noticeable when comparing the count of MA islands of size less than 1.0 $\mu$m$^2$ with the range [1.0, 2.0) $\mu$m$^2$. 
While the population difference between these bins are significant at 400$^o$C it is reduced at 500$^o$C. Refining the ranges of the individual bins in Fig.~\ref{fig:PAM_temp} unravels no disparity in the population distribution of MA islands, as shown in its subplot. 

The spread of MA island counts across the different ranges of area is separated based on composition and shown in Fig.~\ref{fig:PAM_comp}. 
There is noticeable disparity in the densely populated range of upto 2.0 $\mu$m$^2$ between MA islands associated with different steels. 
In type 1 and 11 steels, area of most MA islands fall within 1.0 $\mu$m$^2$, whereas in type 2 and CFB2 steels, maximum count lie between 1.0 and 2.0 $\mu$m$^2$. 
The population of MA islands with area less than 0.9 $\mu$m$^2$ and [1.0, 2.0) $\mu$m$^2$ is almost equal in type 5 and 10 steels. 
In addition to these differences, except for type 1 steel, considerable fraction of MA islands, greater than outliers, exhibit PAM greater than 10 $\mu$m$^2$.
When compared to aspect ratio, the polygon area metric appears to be noticeably influenced by the composition. 

\subsection{Polygon compactness}

\begin{figure}
    \centering
    
    \begin{subfigure}[b]{1.0\textwidth}
        \includegraphics[width=\textwidth]{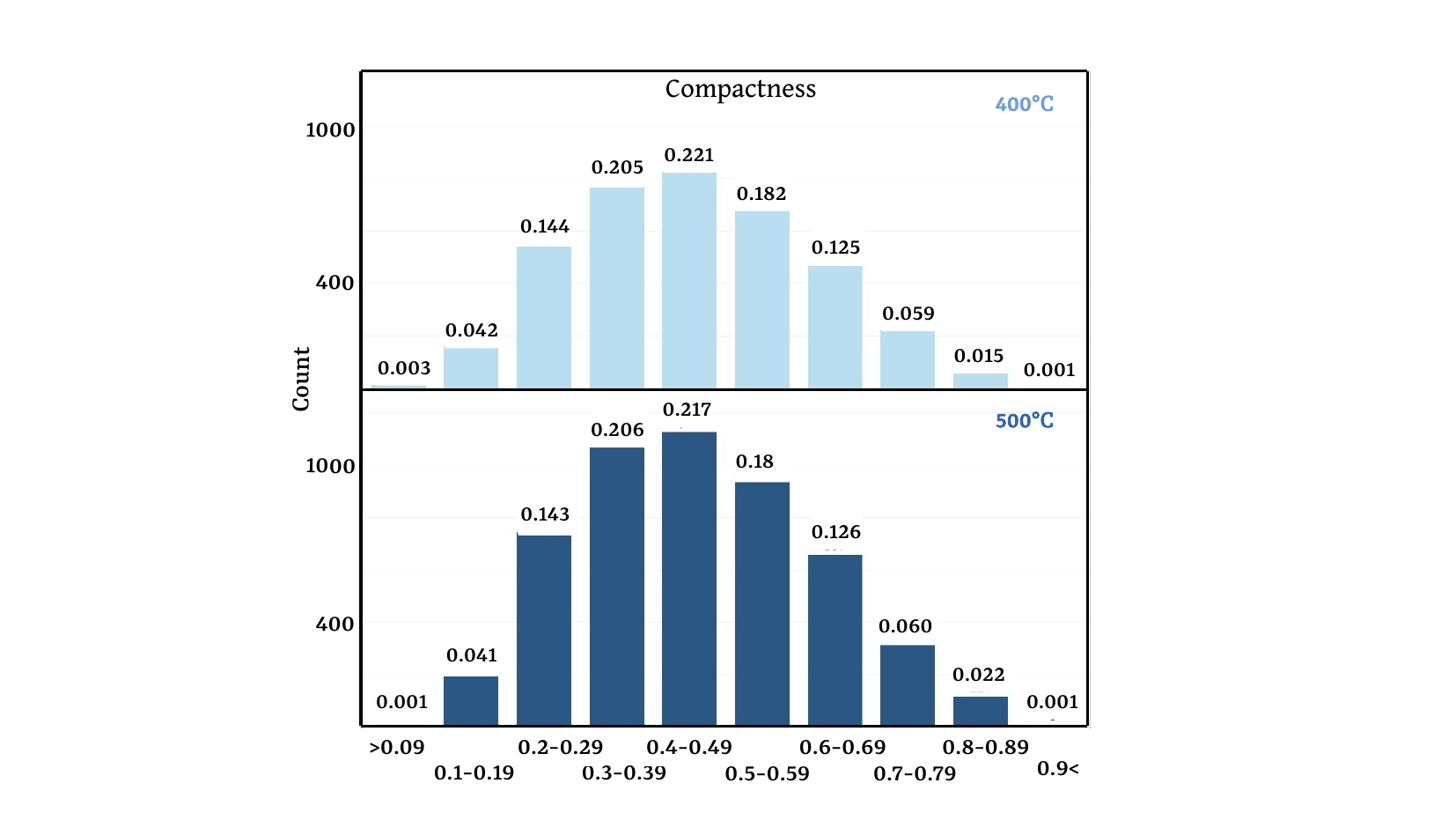}
        \caption{}
        \label{fig:COmpact_temp}
    \end{subfigure}

    \begin{subfigure}[t]{1.0\textwidth}
    \centering
        \includegraphics[width=\textwidth]{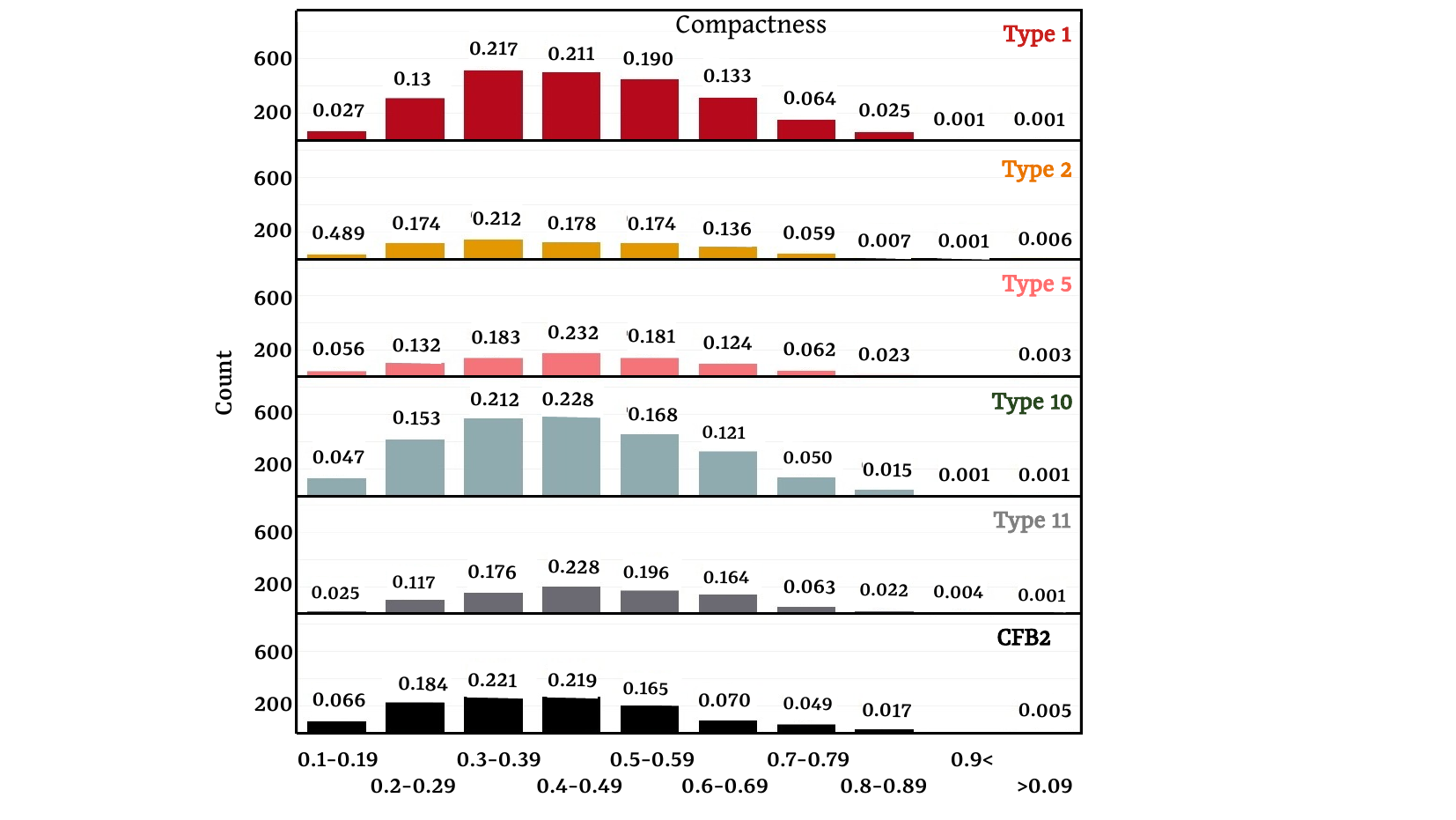}
        \caption{}
        \label{fig:Comp}
    \end{subfigure}
    
    \caption{Polygon compactness based population distribution of MA islands in bainite microstructures (a) transformed at different temperatures and (b) pertaining to different steels. While the counts are represented along the axis, the corresponding number fraction is included within the histogram. }  
    \label{fig:}
\end{figure}

Metadata encompassing geometric information on all MA islands observed in the microstructures includes polygon compactness. 
This feature is proportional to the ratio of area  ($A$) and perimeter ($P$) of MA island  and is generally referred to as Polsby-Popper compactness ($PC=4\pi \frac{A}{P}$).  
Owing to its formulation, polygon compactness of a MA island assumes a value between (0.0, 1.0]. 
Similar to previous representations, population of MA islands are distributed based on compactness,  and is plotted in Fig.~\ref{fig:COmpact_temp}, making a distinction in processing temperature. 
The spread of MA islands across the definite ranges of compactness remains identical, with absolutely no disparity, between the two temperatures. 

Having examined the distribution of polygon compactness across the different temperatures,  the corresponding representation is extended to composition in Fig.~\ref{fig:Comp}. 
Although the overall trend in the distribution of number density is seemingly identical, a closer examination unravels deviations in the densely population sections. 
Compactness between [0.2,0.7) is assumed by most MA islands, irrespective of the composition, with the range [0.3,0.4) or [0.4,0.5) recording relatively highest number-density. 
The disparity in the most populace section of the compactness exist, and they appear rather subtly in the current representation.
In order to explicate the difference in the distribution,  the most populated section of [0.2,0.7) is focused and the average count across this range, for each steel, is individually calculated. 
The population in each bin is now illustrated in view of this average of the focused section. 
To achieve this illustration, a parameter called averaged count, $x_i$ is introduced, which for a given steel is calculated by
\begin{equation}\label{eq:av}
 x_i=\frac{\bar{n}-n_i}{n_i},
\end{equation}
with $\bar{n}$ indicating the average population across the top five densely populated ranging from 0.2 to 0.7 and the count in each of the five bin, [0.2,0.3), [0.3,0.4), [0.4,0.5),[0.5,0.6), and [0.6,0.7), is distinctly represented by $n_i$. 
The parameter $x_i$ can assume a value between $-1.0$ and $1.0$ with negative value indicating a population less than the average of the most populace range and positive, the opposite. 

\begin{figure}
    \centering
      \begin{tabular}{@{}c@{}}
      \includegraphics[width=0.9\textwidth]{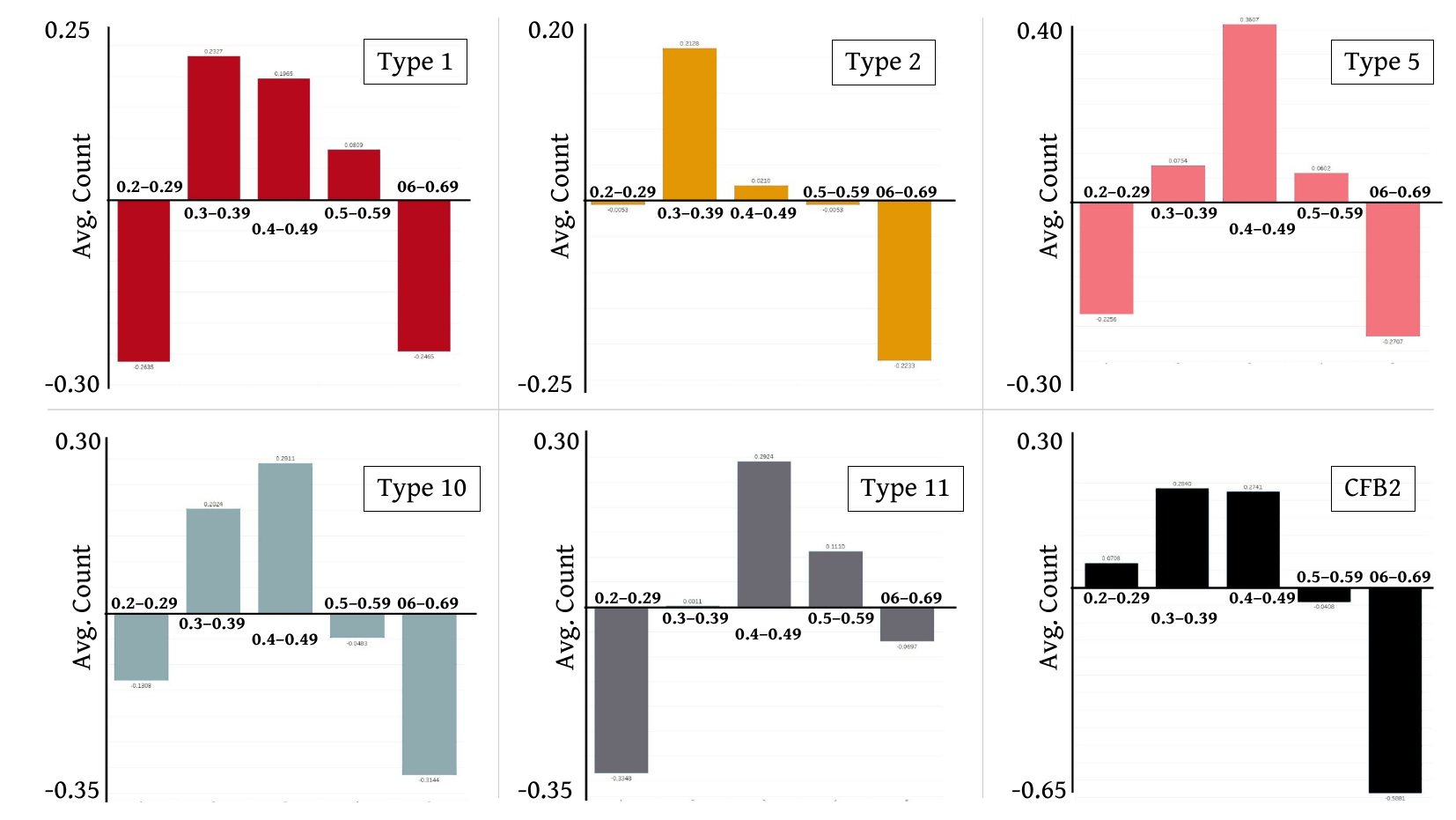}
    \end{tabular}
    \caption{ The distribution of MA island population represented as average count, ascertained through Eqn.~\eqref{eq:av}, across different types of steel. 
    \label{fig:Compact_Avg}}
\end{figure}

The distribution of MA island count based on $x_i$, in the most populated section, across different composition in shown in Fig.~\ref{fig:Compact_Avg}. 
It is evident from this illustration that composition does have noticeable effect on the compactness of MA island. 
While the compactness of most MA island in CFB2, Type 1 and 2 steels range from 0.3 to 0.4, the bin [0.4, 0.5) is most dense in Type 5, 10 and 11 steels. 
Even within each of these two categories of steels the differences are made visible in Fig.~\ref{fig:Compact_Avg}. 
For instance, count in three bins of Type 2 steel is the close to average, $\bar{n}$, with $x_i\approx 0.0$, whereas in Type 1 the population is either noticeably above or below the average. 
Moreover, in type 11, [0.5, 0.6) is second most populated in contrast to type 5 and 10, wherein [0.3, 0.4) assumes corresponding rank.

\section{Conclusion}

Discussions in previous section, and the associated illustrations, indicate few effects of composition and processing temperatures on geometric factors in-keeping with current understanding. 
The relative raise in the count of MA islands with abnormally large area at higher temperature is a prime example. 
These disparities in geometric factors introduced by composition or temperature, however, are often subtle and/or away from the densely populated section of the distribution. 
Moreover, the differences in the predominantly occupied ranges are  largely marginal and need additional refinement or revised representation to explicate them. 
In other words, the general trend in the distributions of MA-island counts across the difference ranges of various geometric factors remain almost the same, irrespective of temperature and composition.
Consequently, instead of a definitive characteristic association, a set of steels can be categorised based on the effect of composition on geometric factors.

\section*{Declaration of Interest}
The authors declare that they have no known competing financial interests or personal relationships that could have appeared to influence the work reported in this paper.

\section*{Acknowledgments}

PGK Amos thanks the financial support of the SCIENCE \& ENGINEERING RESEARCH BOARD (SERB) under the project SRG/2021/000092.

%

\end{document}